\documentclass[aps,pra,twocolumn,a4paper,groupedaddress,floatfix,10pt,longbibliography]{revtex4-1}
\usepackage{graphicx,amsmath,amssymb,amsfonts,dsfont,enumitem,MnSymbol,comment}
\newcommand{\mf}[1]{\boldsymbol{#1}}
\newcommand{\ket}[1]{\ensuremath{|#1\rangle}}
\newcommand{\mc}[1]{\ensuremath{\mathcal{#1}}}
\newcommand{\bra}[1]{\ensuremath{\langle #1 |}}
\newcommand{\braket}[2]{\ensuremath{\langle #1 | #2 \rangle}}

\newcommand{\imag}{\mathrm{i}}
\usepackage{color}

\begin{document}

\title{ 
Mott polaritons in cavity-coupled quantum materials
}
\author{Martin Kiffner${}^{1,2}$}
\author{Jonathan Coulthard${}^{2}$}
\author{Frank Schlawin${}^{2}$}
\author{Arzhang Ardavan${}^{2}$}
\author{Dieter Jaksch${}^{2,1}$}

\affiliation{Centre for Quantum Technologies, National University of Singapore,
3 Science Drive 2, Singapore 117543${}^1$}
\affiliation{Clarendon Laboratory, University of Oxford, Parks Road, Oxford OX1
3PU, United Kingdom${}^2$}

\begin{abstract}
We show that strong electron-electron interactions in 
cavity-coupled quantum materials can enable collectively 
enhanced light-matter interactions with ultrastrong effective 
coupling strengths. 
As a paradigmatic example we 
consider a Fermi-Hubbard model coupled to a single-mode cavity   
and find that resonant electron-cavity interactions result in the formation of a quasi-continuum of  polariton branches. The vacuum Rabi splitting 
of the two outermost branches  is collectively enhanced and scales with $g_{\text{eff}}\propto\sqrt{2L}$, where $L$ is the number of electronic sites, and the maximal achievable value for $g_{\text{eff}}$ is 
determined by the volume of the unit cell of the crystal. We find that $g_{\text{eff}}$ for 
existing quantum materials  can by far exceed the width of the first excited Hubbard band. This effect can be experimentally observed via measurements of the  optical conductivity and does not require ultra-strong coupling on the single-electron level. 
Quantum correlations in the electronic ground state as well 
as the microscopic nature of the light-matter interaction  
enhance the collective light-matter interaction 
compared to an ensemble of independent two-level atoms interacting with a cavity mode. 
\end{abstract}
 
\maketitle

\section{Introduction \label{intro}}
Collective phenomena in light-matter interactions are of 
tremendous interest in quantum physics. The characteristic 
feature of these phenomena is that observable quantities increase with the number of emitters, and thus intrinsically small quantum effects 
can be elevated to a macroscopic level.  
One of the first studied examples 
is superradiance within the Dicke model~\cite{dicke2:54,garraway:11}, 
which comprises an ensemble of independent  two-level atoms 
interacting with a single mode of the radiation field.

Collective light-matter interactions are conveniently described 
within the framework of polaritons, which are combined excitations 
of light and matter. A prominent example is given by dark-state polaritons in laser-driven atomic gases~\cite{fleischhauer:05} and more 
recently,  polaritons  have been investigated in 
various solid state systems coupled to  
cavities~\cite{hui:10,carusotto:13,orgiu:15,schwartz:11,kenacohen:13,zhang:14,tabuchi:14,yao:15,sivarajah:17b,abdurakhimov:18,mergenthaler:17,hagenmueller:10,scalari:12,zhang:16,li:18,paravicini:18,bartolo:18}.  For example, Bose-Einstein condensation of exciton polaritons in semiconductor  materials attracted considerable 
attention~\cite{hui:10,carusotto:13}, 
and molecular systems~\cite{orgiu:15,schwartz:11,kenacohen:13} 
coupled to cavities can exhibit giant Rabi splittings between polariton branches.
The strong coupling of  magnetic excitations to microwave 
cavities was investigated in ~\cite{zhang:14,tabuchi:14,yao:15,sivarajah:17b,abdurakhimov:18,mergenthaler:17}, and 
 two-dimensional electron gases coupled to THz cavities were 
 studied in~\cite{hagenmueller:10,scalari:12,zhang:16,li:18,paravicini:18,bartolo:18}. In all these systems~\cite{hui:10,carusotto:13,orgiu:15,schwartz:11,kenacohen:13,zhang:14,tabuchi:14,yao:15,sivarajah:17b,abdurakhimov:18,mergenthaler:17,hagenmueller:10,scalari:12,zhang:16,li:18,paravicini:18,bartolo:18}, Coulomb interactions between electrons play a minor role and are not 
 directly involved in the formation of polaritons. 
A particularly intriguing yet challenging platform for investigating light-matter 
interactions are quantum materials~\cite{editorial:16,powell:06,powell:11,kato:04}. 
In these systems strong electron-electron interactions give rise a 
plethora of physical effects that are difficult to describe 
due to their intrinsically quantum many-body nature. 
An example is given by the Mott metal-insulator transition~\cite{mott:49,masatoshi:98} 
which can be modelled within the Fermi-Hubbard 
model~\cite{essler:05}. 

First steps investigating how quantum materials couple to classical and quantum light have been undertaken recently. 
The interaction of quantum materials  with  strong, classical 
light fields was investigated in~\cite{mentink:15,coulthard:17,goerg:18,stepanov:17}, and superradiance of quantum materials coupled to a cavity field was predicted in~\cite{mazza:18}. 
The possibility of  inducing superconductivity 
by coupling electron systems to terahertz and microwave cavities was 
explored in~\cite{laplace:16,schlawin:18,curtis:18,sentef:18}. 
Furthermore, it was shown in~\cite{kiffner:19,kiffner:19b} that  
second-order electron-cavity interactions  reduce the 
magnetic exchange energy in cavity-coupled quantum materials and lead to 
a collectively enhanced momentum-space pairing effect for 
electrons. 
%
\begin{figure}[t!]
\begin{center}
\includegraphics[width=\columnwidth]{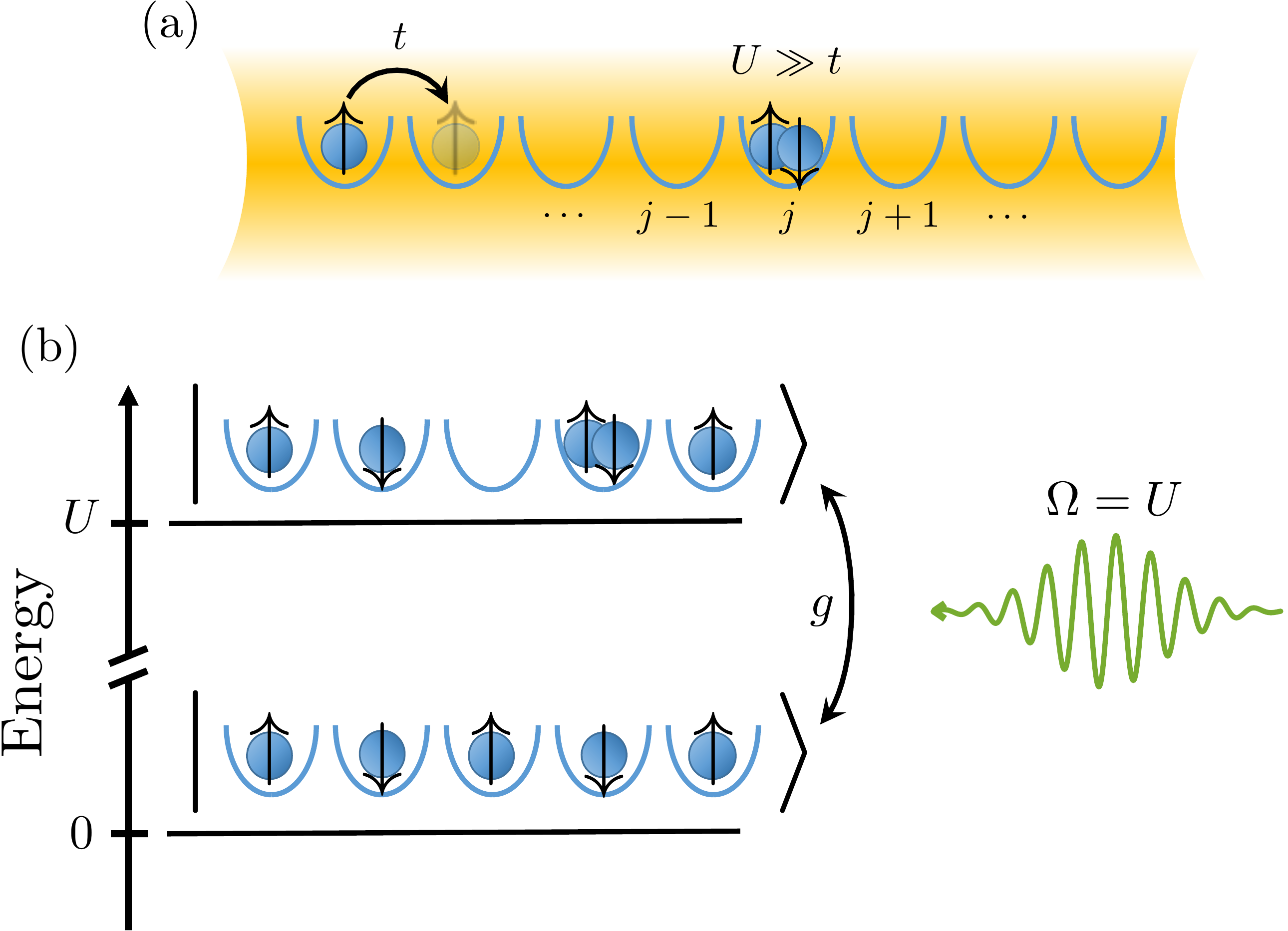}
\end{center} 
\caption{\label{fig1}
(Color online) (a) The system of interest comprises an electronic 
system that is weakly coupled to a single-mode cavity 
with resonance frequency $\omega_c$ and photon energy $\Omega = \hbar \omega_c$. We consider a one-dimensional Fermi-Hubbard model with on-site interaction $U$ and hopping amplitude $t$ for the 
electronic system. 
(b) Schematic illustration of the resonant electron-photon coupling 
with coupling strength $g$ between 
the electronic ground state and a doublon-holon pair state with energy $U$. 
}
\end{figure}
%

Here we show that  strong electron-electron interactions in 
cavity-coupled quantum materials can enable collectively 
enhanced light-matter interactions that change the macroscopic 
properties of the quantum material. 
In particular, we consider a one-dimensional Hubbard model coupled to a single-mode cavity as shown in Fig.~\ref{fig1}(a). We find that the optical conductivity of the quantum material features two peaks that are separated in energy by the collectively enhanced vacuum Rabi frequency $g_{\text{eff}}\propto \sqrt{2L}$, where $L$ is the number of electronic sites.   
Macroscopically large energy splittings are thus even 
possible for weakly 
coupled electron-photon systems. The largest possible value of 
$g_{\text{eff}}$ is attained if the material fills the entire cavity. 
In this case, the effective coupling constant becomes independent 
of $L$ and $g_{\text{eff}}\propto 1/\sqrt{v_{\text{uc}}}$, where $v_{\text{uc}}$ is the volume of the unit cell of the crystal.
As an example, for the quantum material $\text{ET-F}_2\text{TCNQ}$,  which is well described by a one-dimensional Hubbard model, $g_{\text{eff}}$ can exceed $250 \text{meV}$. This  is 
several orders of magnitude larger than collective energy shifts in 
atomic systems~\cite{raizen:89,thompson:92,baumann:10} and comparable 
to the extremely large energy splittings achieved in cavity-coupled molecular materials~\cite{orgiu:15,schwartz:11,kenacohen:13}. 

The  resonant light-matter interactions 
considered here are schematically shown in Fig.~\ref{fig1}(b). 
An electronic state at half filling and with no electronic excitation is 
resonantly coupled via a cavity photon to an electronic state with one 
doubly occupied state (doublon) and an empty site (holon) next to it. 
These states differ in energy by the on-site Coulomb interaction $U$, 
which corresponds to the Mott gap of the quantum material.   
The transition dipole moment between these two states is of 
the order of $d e$ ($d:$  lattice spacing,  $e$ elementary charge), 
which is comparable to strong transitions in alkali metal atoms~\cite{steck}. 
A single-mode cavity is tuned in resonance with this transition 
between electronic states. 

We find that this resonant electron-photon interaction leads to 
a quasi-continuum of polariton states. The two branches with the 
largest energy splitting $g_{\text{eff}}$ can be constructed from 
the electronic ground state. The collective energy splitting  
$g_{\text{eff}}\propto \sqrt{2L}$ of these branches 
gives rise to the two peaks in the optical conductivity. 
A comparison of our results with the Dicke 
model~\cite{dicke2:54,tavis:68,tavis:69,garraway:11}
reveals two important differences between the two systems. 
First, the quantum correlations in the ground state of the electronic 
system lead to an enhancement of $g_{\text{eff}}$ by $\approx18\%$. 
Second, $g_{\text{eff}}$ for the electronic system 
is larger by a factor of $\sqrt{2}$ than in the Dicke model. We show that this difference is caused by the different microscopic nature of the light-matter coupling in these systems.

This paper is organized as follows. The theoretical model describing the system shown in Fig.~\ref{fig1}(a) is introduced in Sec.~\ref{model}. 
Our results are presented in Sec.~\ref{results}, and the derivation 
of the Mott polaritons in the manifold with one excitation is outlined 
in Sec.~\ref{Mpol}. We then show in  Sec.~\ref{optcon} that a direct 
signature of the light-matter hybridisation appears in the optical conductivity. 
The discussion in Sec.~\ref{discuss} gives an intuitive explanation 
 for the collective enhancement of the polariton splitting and illustrates similarities and differences of our system with the Dicke model. 
The experimental observation of the predicted effects is 
discussed in Sec.~\ref{exp}, 
and a summary of our results is provided in Sec.~\ref{summary}. 
\section{Model \label{model}}
 In this section we present the 
theoretical model established in~\cite{kiffner:19,kiffner:19b} 
for the quantum hybrid system shown in Fig.~\ref{fig1}(a). 
The electronic system is described by the one-dimensional Fermi-Hubbard model~\cite{essler:05} with on-site energy $U$ and 
hopping amplitude $t$. The electrons are weakly coupled 
to a single-mode cavity with resonance frequency $\omega_c$, 
and  $\Omega =\hbar\omega_c$ is the photon energy.

\begin{figure}[t!]
\begin{center}
\includegraphics[width=\columnwidth]{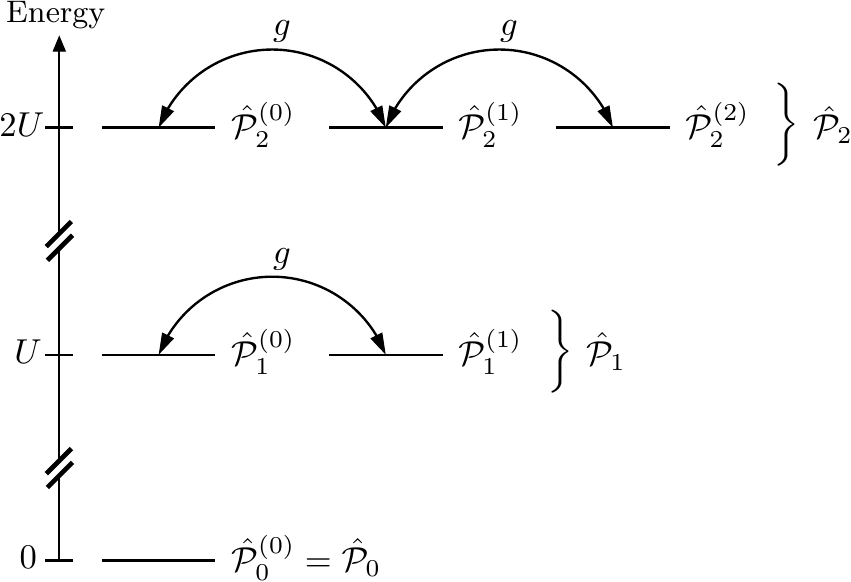}
\end{center}
\caption{\label{fig2}
%
Schematic drawing of the spectrum of $\hat{H}_0 = \hat{D}  + \hat{P}$ 
for $U=\Omega$.  $\hat{\mc{P}}_n^{(j)}$ projects onto  
a sub-manifold with $j$ photons and $n-j$ doublons, and $\hat{\mc{P}}_n$ is 
the projector onto the manifold with $n$ excitations. Only manifolds with $n\le 2$ excitations are shown. The 
resonant cavity coupling $g$ between sub-manifolds is indicated by arrows. Higher-order couplings induced by $\hat{V}$ are not shown.
}
\end{figure}

The gross energy structure of our system in the parameter regime 
of interest ($U,\, \Omega\gg t$) is determined by the Hamiltonian 
\begin{align}
\hat{H}_0  = \hat{P} + \hat{D} \,, 
 \label{h0}
\end{align}
where 
\begin{align}
\hat{P} = \Omega \hat{a}^{\dagger} \hat{a}
\label{Hphot}
\end{align}
describes the cavity photons  and 
$\hat{a}^{\dagger}$ ($\hat{a}$) is the bosonic photon creation 
(annihilation) operator. 
The operator $\hat{D}$ in Eq.~(\ref{h0})  accounts for  the on-site Coulomb repulsion between electrons, 
 \begin{align}
 \hat{D} =& U \sum\limits_{k=0}^{L} k\, \hat{\mc{P}}_k^{D} \,,
 \label{Ds}
 \end{align}
where $U$ is the  interaction energy and $\hat{\mc{P}}_k^{D}$ is the projector 
onto the manifold with $k$ doubly-occupied sites~\cite{kiffner:19}. 
In the following we 
refer to these excitations as doublons.  The  eigenstates  of $\hat{H}_0$ are tensor products of photon number states $\ket{j_P}$ with  $j$ photons and Wannier states~\cite{essler:05} with $k$ doublons. The associated eigenvalues  $j \Omega +k U$ are generally highly  degenerate and form manifolds as shown in Fig.~\ref{fig2}. 
We denote the projector onto a manifold with $j$ photons and $k=n-j$ 
doublons by 
\begin{align}
 \hat{\mc{P}}_n^{(j)} = \hat{\mc{P}}_{n -j }^{D}\otimes \hat{\mc{P}}_{j}^{P} \,,
 \label{pnj}
\end{align}
where $n$ is the  total number of excitations, 
$\hat{\mc{P}}_{j}^{P}=\ket{j_{P}}\bra{j_{P}}$ projects onto the 
subspace with $j$ photons and 
\begin{align}
\hat{\mc{P}}_n =  \sum_{j=0}^{n} \hat{\mc{P}}_n^{(j)}
 \label{pn}
\end{align}
projects onto all  sub-manifolds with $n$ excitations. 

Modifications to the simple energy structure shown in Fig.~\ref{fig1} 
arise from the electron hopping and the electron-photon interaction. 
The hopping operator is 
\begin{align}
  \hat{T} = & -t \sum\limits_{ \langle jk\rangle \sigma} \left(\hat{c}_{j,\sigma}^{\dagger}\hat{c}_{k,\sigma} 
 + \text{h.c.}\right)  \,,
 \label{T} 
\end{align}
where  $\langle j k\rangle$ denotes neighbouring sites with $j<k$ 
 and $\hat{c}_{j,\sigma}^{\dagger}$ ($\hat{c}_{j,\sigma}$) creates (annihilates) an electron at 
 site $j$ in spin state $\sigma\in\{\uparrow,\downarrow\}$. 

The electron-photon interaction was derived 
in~\cite{kiffner:19,kiffner:19b} via the Peierls 
substitution~\cite{essler:05} and by expanding the resulting 
interaction Hamiltonian up to second order in the electron-cavity coupling, 
\begin{align}
 \hat{V}=  
 g (\hat{a}+\hat{a}^{\dagger}) \hat{\mc{J}} 
 -\frac{1}{2}\frac{g^2}{t^2}(\hat{a}+\hat{a}^{\dagger})^2\hat{T}\,,
\label{Vint}
\end{align}
where 
 \begin{align}
 \hat{\mc{J}} = &-\imag  \sum\limits_{\langle j k \rangle \sigma } \left(\hat{c}_{j,\sigma}^{\dagger}\hat{c}_{k,\sigma} 
 -\hat{c}_{k,\sigma}^{\dagger}\hat{c}_{j,\sigma} \right) 
 \label{current}
\end{align}
is the  dimensionless current operator. 
The parameter $g=t \eta$ in $\hat{V}$ determines the coupling strength between the electrons and photons,  and
\begin{align}
 \eta=\frac{d e}{\sqrt{2\hbar\varepsilon_0 \omega_c v}}
 \label{eta}
\end{align}  
is a dimensionless parameter that 
depends on the lattice constant $d$ and  the cavity  mode volume $v$ 
($e$: elementary charge, $\varepsilon_0$: vacuum permittivity, $\hbar$: reduced Planck's constant). 
The derivation of $\hat{V}$  assumes~$\eta \ll 1$, and this condition 
is also required to grant the validity of the single-mode cavity approximation~\cite{munoz:18}.
%

%
%
With the preceding definitions we arrive at the total Hamiltonian for the quantum hybrid system in Fig.~\ref{fig1}, 
\begin{align}
 %
\hat{H}  = \hat{H}_0 + \hat{H}_1\,,
%
 %
\label{totv}
\end{align}
where 
\begin{subequations}
\begin{align}
 \hat{H}_1 & =  \hat{T}+\hat{V}\,, \\
& = \left[1-\frac{1}{2}\frac{g^2}{t^2}(\hat{a}+\hat{a}^{\dagger})^2\right]\hat{T} + 
 g (\hat{a}+\hat{a}^{\dagger}) \hat{\mc{J}} \,.
\end{align}
\end{subequations}
For the parameters of interest ($U, \Omega\gg t\gg g$), $H_1$ can be treated 
as a perturbation to the gross energy structure dictated by $H_0$. 

In the following Sec.~\ref{results} we investigate the formation of 
Mott polaritons through resonant electron-photon interactions in 
$\hat{\mc{P}}_1$. To set the stage for this we recall 
how the cavity modifies the physics in $\hat{\mc{P}}_0$, which 
was investigated in~\cite{kiffner:19,kiffner:19b} using second-order 
perturbation theory. 
For the special case of $\Omega=U$ and  an electronic system at half filling, the effective Hamiltonian in $\hat{\mc{P}}_0$ is given 
by~\cite{kiffner:19,kiffner:19b}
\begin{align}
  \hat{H}_{\text{eff}}^{(0)} = \hat{H}_{\text{S}}  \otimes\hat{\mc{P}}_0^P\,, 
 \label{resg}
\end{align}
where 
\begin{align}
\hat{H}_{\text{S}} = & -J_c \hat{\mc{P}}_0^D\left(
\sum\limits_{\langle kl \rangle} \hat{b}_{kl}^{\dagger}\hat{b}_{kl}\right)
\hat{\mc{P}}_0^D \label{Hex} 
\end{align}
and 
 \begin{align}
 \hat{b}_{kl}^{\dagger} = \left( \hat{c}_{k,\uparrow}^{\dagger} \hat{c}_{l,\downarrow}^{\dagger} 
 - \hat{c}_{k,\downarrow}^{\dagger} \hat{c}_{l,\uparrow}^{\dagger}\right)/\sqrt{2} 
\label{bkl}
\end{align}
creates a singlet pair at sites $k$ and $l$. 
 $\hat{H}_{\text{S}}$ acts only on the electronic 
system and is an  isotropic Heisenberg model [see Appendix~\ref{hbc}] 
with coupling 
\begin{align}
 J_c  & =\mc{R}_c \frac{4  t^2}{U}\,, \label{Jc}
\end{align}
where 
\begin{align}
 \mc{R}_c  & = 1 -\frac{1}{2}\frac{g^2}{t^2} \label{Rc}
\end{align}
is a dimensionless scaling factor. Note that $R_c$ is 
equal to unity for $g=0$ and $R_c<1$ for $g>0$, and thus 
the cavity reduces the magnetic exchange interaction. In addition, we have $J_c>0$ for all permitted values of $g\ll t$ and thus the ground state 
$\ket{G}$ of 
$\hat{H}_{\text{S}}$ is  an antiferromagnetic state~\cite{essler:05}. 
\section{Results \label{results}} 
Throughout this section we consider an electronic system at half filling 
and $\Omega=U$. In Sec.~\ref{Mpol} we show that resonant electron-photon 
interactions result in the formation of  polaritons, and the 
energy splitting of the two outermost polariton branches 
is collectively enhanced. 
Evidence for this light-matter hybridisation 
can be found in the optical conductivity as shown in Sec.~\ref{optcon}. 
\subsection{Mott polaritons \label{Mpol}}
The  first excited manifold $\hat{\mc{P}}_1$  contains all states with either one doublon or one photon. 
The effective Hamiltonian in $\hat{\mc{P}}_1$ and  in first order in $H_1$ 
is~[see Appendix~\ref{sx}]
\begin{align}
  \hat{H}_{\text{eff}}^{(1)}= U \hat{\mc{P}}_1 + \mc{R}_c\hat{\mc{P}}_1^{(0)}\hat{T}\hat{\mc{P}}_1^{(0)}
   + \hat{H}_{D-P} \,.
 \label{oneEx}
\end{align}
Higher-order terms in $H_1$ are neglected in Eq.~(\ref{oneEx})  
and become negligible in the limit $U\gg t,g$. 
The first term in Eq.~(\ref{oneEx}) is a constant energy 
offset of the states in $\hat{\mc{P}}_1$.  The second term 
describes the dynamics of the doublon and holon in $\hat{\mc{P}}_1$ and  
gives rise to the first excited Hubbard band. At $g=0$, 
the width of this band is $8 t$~\cite{gallagher:97,jeckelmann:03}, and 
the scaling factor $\mc{R}_c$ reduces this width slightly for $g>0$. 
The last term in Eq.~(\ref{oneEx}) accounts for the resonant doublon-photon interaction and is given by~[see Appendix~\ref{SpecHdp}], 
\begin{align}
 \hat{H}_{D-P}  = g \left( \hat{\mc{D}}^{\dagger}\otimes\hat{\mc{A}} + \hat{\mc{D}}\otimes\hat{\mc{A}}^{\dagger} \right) \,, 
 \label{Hdp}
\end{align}
where 
 \begin{align}
 \hat{\mc{A}}= \ket{0_P}\bra{1_P} 
 \end{align}
is a transition operator between the vacuum and the one-photon state and 
\begin{align}
 \hat{\mc{D}} = \hat{\mc{P}}_0^{D}\hat{\mc{J}}\hat{\mc{P}}_1^{D}
 \label{Dop}
\end{align}
mediates a transition between one and  zero doublon states. 
We note that the definition of $\hat{\mc{J}}$ allows us 
to write $\hat{\mc{D}}$ as 
\begin{align}
  \hat{\mc{D}} = -\imag \left(\hat{\mc{D}}_R - \hat{\mc{D}}_L \right) \,,
 \label{pmd}
\end{align}
where 
\begin{subequations}
 \begin{align}
  \hat{\mc{D}}_R = & \hat{\mc{P}}_0^{D}
  \left(\sum\limits_{\langle j k \rangle \sigma } \hat{c}_{j,\sigma}^{\dagger}\hat{c}_{k,\sigma}\right) \hat{\mc{P}}_1^{D}\\
\hat{\mc{D}}_L = & \hat{\mc{P}}_0^{D}
\left(\sum\limits_{\langle j k \rangle \sigma }
\hat{c}_{k,\sigma}^{\dagger}\hat{c}_{j,\sigma} \right) 
\hat{\mc{P}}_1^{D}\,.
 \end{align}
\end{subequations}
Since $j<k$ in $\langle j k\rangle$, this means that $\hat{\mc{D}}_R$ 
$(\hat{\mc{D}}_R^{\dagger})$ annihilates (creates) a doublon-holon pair 
where the doublon is to the right of the holon. Similarly, 
$\hat{\mc{D}}_L$ 
$(\hat{\mc{D}}_L^{\dagger})$ annihilates (creates) a doublon-holon pair 
where the doublon is to the left of the holon.

We emphasize that $\hat{H}_{D-P}$ is of first order in the  electron-photon coupling since it is proportional to the coupling strength $g$. 
This is in contrast to the ground-state manifold  where the leading term 
is of second order in the electron-photon coupling~\cite{kiffner:19,kiffner:19b}. We thus expect that the electron-cavity coupling 
has a much stronger effect in $\hat{\mc{P}}_1$ than in $\hat{\mc{P}}_0$ for a fixed value of $g$. 
The resonant electron-photon coupling described by $\hat{H}_{D-P}$ in Eq.~(\ref{Hdp}) results 
in the formation of doublon-photon polaritons. 
All eigenstates of $\hat{H}_{D-P}$ with non-zero eigenvalues can be constructed from the eigenstates  
of $\hat{H}_{\text{S}}$ with non-zero eigenvalues~[see Appendix~\ref{SpecHdp}]. 
For each eigenstate $\ket{g_j}$ with 
\begin{align}
\hat{H}_{\text{S}} \ket{g_j} =\mc{E}_j \ket{g_j}
\end{align}
and $\mc{E}_j<0$ the corresponding pair of polariton states is 
 \begin{align}
 \ket{\psi_{D-P}^{j}}_{\pm}  = \frac{1}{\sqrt{2}}\left[\ket{g_j}\otimes\ket{1_P} \pm   \frac{1}{2\epsilon_j} \left(\hat{\mc{D}}^{\dagger}\ket{g_j}\right)\otimes\ket{0_P}\right] \,,
  \label{polariton}
 \end{align}
where 
\begin{align}
\epsilon_j=\sqrt{-\mc{E}_j/J_c}  
\label{epsj}
\end{align}
and
 \begin{align}
 \hat{H}_{D-P} \ket{\psi_{D-P}^{(j)}}_{\pm}  = \pm 2g \epsilon_j \ket{\psi_{D-P}^{(j)}}_{\pm}. 
 \end{align}
Each polariton state in Eq.~(\ref{polariton}) is a maximally entangled superposition of a state with 
one doublon and no photon, and a 
state with no doublon and one photon.  The largest value 
$\epsilon_{\text{max}}$ 
corresponds to the ground state $\ket{G}$ of $\hat{H}_{\text{S}}$ in 
Eq.~(\ref{resg}) with energy $\mc{E}_G$, and 
thus
\begin{align}
 \epsilon_{\text{max}} =\sqrt{-\mc{E}_G/J_c}\approx\sqrt{L\log 2}
\end{align}
for $L\gg 1$~\cite{mossel:08}. The energy difference between the corresponding polariton 
states  $\ket{\psi_{D-P}^{\text{max}}}_{\pm}$ 
is  
\begin{align}
g_{\text{eff}}=4 g\sqrt{L\log 2}\approx 3.33 g \sqrt{L}, 
\label{geff}
\end{align}
and carries a direct signature of the collective doublon-photon coupling. 
If the material fills the mode volume $v$ of the cavity we have 
$v=L v_{\text{uc}}$, where $v_{\text{uc}}$ is  the volume per lattice site. Since 
$g = \eta t \propto 1/\sqrt{v}$ [see Eq.~(\ref{eta})], 
  the value of $g_{\text{eff}}$ is independent of $L$ and 
just depends on $v_{\text{uc}}$. 
This shows that nanoplasmonic cavities are  not required to achieve large values of $g_{\text{eff}}$.

The  values of  $\epsilon_j$ for a system with $L=12$ sites are shown in 
Fig.~\ref{fig4}(a), and the corresponding density of states is shown in Fig.~\ref{fig4}(b). 
Even a relatively small system 
with $L=12$ sites exhibits a quasi-continuum of polariton states, with the largest density of 
states for intermediate values of $\epsilon_j$. 
%
%
\begin{figure}[t!]
\begin{center}
\includegraphics[width=\columnwidth]{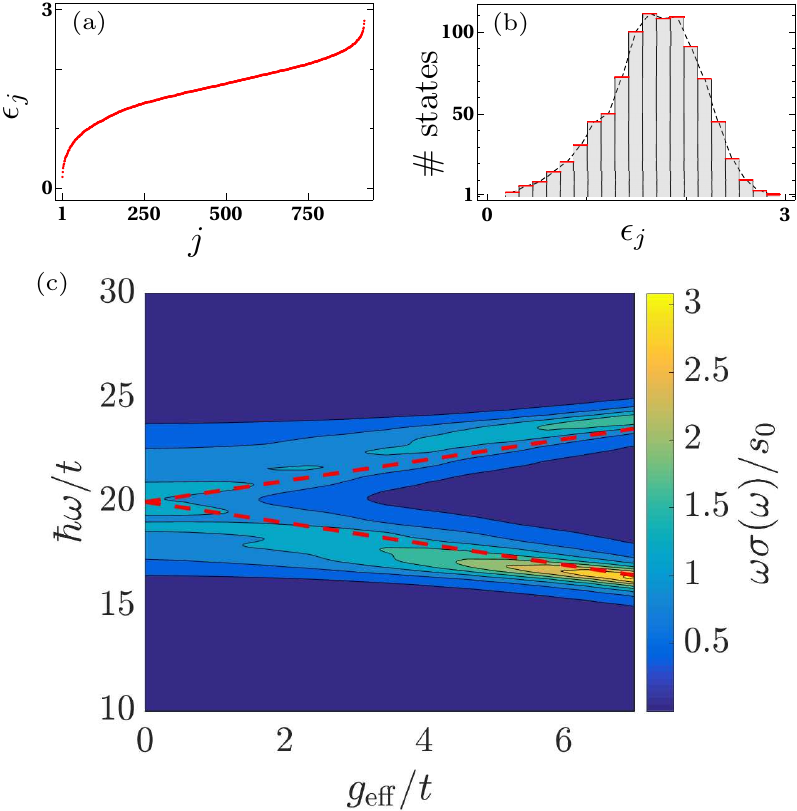}
\end{center}
\caption{\label{fig3}
(Color online) (a) All values of $\epsilon_j$ for a system 
with $L=12$ sites at half filling and total spin $S=0$. 
(b) Density of states of the $\epsilon_j$ values shown in (a). 
(c) Density plot of $\omega\sigma(\omega)$ for $L=12$ sites at half filling, zero temperature and $U=\Omega=20 t$, 
where $\sigma(\omega)$ is the optical conductivity and $s_0=t e^2/(\hbar^2 d)$. 
$g_{\text{eff}}=3.33 g \sqrt{L}$ is the collective coupling strength, and 
the red dashed lines indicate the energies $\pm 0.5 g_{\text{eff}}$ of $\ket{\psi_{D-P}^{\text{max}}}_{\pm}$. 
Each term of the sum in Eq.~(\ref{OCsimple})  
was artificially broadened with a Lorentzian of width $0.5 t$.
}
\end{figure}
%

Note that the states $\ket{\psi_{D-P}^{j}}_{\pm}$ in 
Eq.~(\ref{polariton}) are not eigenstates of the full effective Hamiltonian in Eq.~(\ref{oneEx}) due to the kinetic 
energy term 
$\hat{\mc{P}}_1^{(0)}\hat{T}\hat{\mc{P}}_1^{(0)}$. 
However, we show in Appendix~\ref{SpecHdp} that the states 
$\ket{\psi_{D-P}^{j}}_{\pm}$ are approximate eigenstates of 
$\hat{H}_{\text{eff}}^{(1)}$ if their energy splitting is 
much larger than $t$. In this case, $\hat{\mc{P}}_1^{(0)}\hat{T}\hat{\mc{P}}_1^{(0)}$ only leads to a  broadening 
of the polariton states by coupling them off-resonantly to the 
quasi-continuum of the first Hubbard band. 
\subsection{Optical conductivity \label{optcon}}
A direct signature of the collective doublon-photon coupling in the first excited manifold can 
be found in the  optical conductivity~\cite{essler:05}, 
\begin{align}
 \sigma(\omega) = \frac{\pi  e^2 t^2}{d \hbar^3} \sum\limits_{m>0}\frac{|\bra{\psi_m} 
 \hat{\mc{J}}\ket{\psi_0}|^2}{\omega_m-\omega_0}\delta[\omega-(\omega_m-\omega_0)]\,,
 \label{OCsimple}
\end{align}
where $\ket{\psi_m}$ are the eigenstates of the full Hamiltonian $\hat{H}$ with energies $E_m=\hbar\omega_m$ 
and $E_0=\hbar\omega_0$ is the energy of the ground state $\ket{\psi_0}$.  
We calculate the optical conductivity with the full system Hamiltonian using Krylov subspace 
methods~\cite{hochbruck:11} for a half-filled electronic system with $L=12$ sites. 
Figure~\ref{fig4}(c) shows  a density  plot of the optical conductivity spectrum as a 
function of  $g_{\text{eff}}$ and $\omega$. 
At $g_{\text{eff}}=0$ the optical conductivity maps out the first excited Hubbard band of 
width $8t$ that describes the kinematic excitations of  a single doublon.  
At $g_{\text{eff}}/t\approx 3$ the optical conductivity splits into two branches that become narrower with 
increasing $g_{\text{eff}}$. The peaks of the optical conductivity signal 
approximately follow the energies of 
the  polariton branches $\ket{\psi_{D-P}^{\text{max}}}_{\pm}$. 

These results suggest that the optical conductivity signal is mostly dominated by 
the two outermost polariton branches $\ket{\psi_{D-P}^{\text{max}}}_{\pm}$ for $g_{\text{eff}}/t\ge 3$ 
which can be understood as follows. 
States which are split strongly by the cavity's light field are also expected to couple 
strongly to an externally applied light field, and  thus they show a strong signal in the optical conductivity. 
This can also be confirmed by noting that 
$\ket{\psi_m}=\ket{\psi_{D-P}^{\text{max}}}_{\pm}$ are the only polariton states contributing to the 
sum in Eq.~(\ref{OCsimple}) if we  approximate the ground state by $\ket{\psi_0}\approx \ket{G}\otimes\ket{0_P}$.  
The finite width of the optical conductivity signal is caused by the coupling of polariton states to 
the first Hubbard band via $\hat{\mc{P}}_1^{(0)}\hat{T}\hat{\mc{P}}_1^{(0)}$, 
and this coupling becomes less effective with increasing energy splitting  $\propto g_{\text{eff}}$~[see Appendix~\ref{SpecHdp}]. 
The slight asymmetry in the intensity and 
position of the two conductivity branches as well as the  
slight increase of the energy splitting compared with the analytical result 
is a consequence of  the higher order terms that are neglected 
in Eq.~(\ref{oneEx}) but taken into account  
in the numerical evaluation of $\sigma(\omega)$. 
\section{Discussion \label{discuss}}
In Sec.~\ref{results} we have shown that there is a one-to-one correspondence between the polariton branches in the manifold 
$\hat{\mc{P}}_1$ and the eigenstates of  $\hat{H}_{\text{S}}$ 
in Eq.~(\ref{Hex}) with zero excitations. The two branches with the largest splitting $g_{\text{eff}}$ contribute significantly to the optical conductivity signal and correspond 
to the electronic ground state $\ket{G}$ of $\hat{H}_{\text{S}}$. 
A rigorous derivation of the results presented in Sec.~\ref{results} 
is provided in  Appendix~\ref{SpecHdp}. 
Here we give an alternative and approximate derivation of the 
two polariton branches with the largest energy splitting $g_{\text{eff}}$. 
This more intuitive picture allows us to gain further insights into 
our system and highlights similarities and 
differences with other polariton systems. 

Our elementary derivation of $g_{\text{eff}}$ starts by approximating 
the electronic ground state $\ket{G}$ of $\hat{H}_{\text{S}}$ by 
$\ket{G}\approx \ket{G_{\text{N\'{e}el}}}$, where 
\begin{align}
 \ket{G_{\text{N\'{e}el}}} = \ket{\uparrow_1,\downarrow_2,\uparrow_3,\ldots}
\end{align}
is the antiferromagnetic N\'{e}el state. Note that we also employed this 
state for $L=5$ to illustrate the electronic ground state in 
Fig.~\ref{fig1}(b). 
Applying the doublon-holon creation operator 
$\hat{\mc{D}}^{\dagger}$ [see Eq.~(\ref{Dop})] to this state 
results in a state with one doublon excitation, 
\begin{align}
 \ket{E_{\text{N\'{e}el}}} = \imag\, \mc{C} 
 \left(\hat{\mc{D}}_R^{\dagger} - \hat{\mc{D}}_L^{\dagger} \right)
 \ket{G_{\text{N\'{e}el}}}\,,
\end{align}
where $\mc{C}$ is a normalisation constant. Assuming open boundary 
conditions, the operators  
$\hat{\mc{D}}_L^{\dagger}$ and $\hat{\mc{D}}_R^{\dagger}$  each  
create $L-1$ states with a holon-doublon pair. Since all these states are orthonormal, we have 
$\mc{C}=1/\sqrt{2(L-1)}\approx1/\sqrt{2L}$ for $L\gg1$. Ignoring boundary 
effects the matrix element of $\hat{H}_{D-P}$ between the states 
$\ket{G_{\text{N\'{e}el}}}\otimes\ket{1_P}$ and 
$\ket{E_{\text{N\'{e}el}}}\otimes\ket{0_P}$ is thus
\begin{align}
 \left[\bra{E_{\text{N\'{e}el}}}\otimes\bra{0_P}\right] 
 \hat{H}_{D-P}
 \left[\ket{G_{\text{N\'{e}el}}}\otimes\ket{1_P}\right] 
 \approx  g \sqrt{2 L}\,.
\end{align}
Diagonalisation of $\hat{H}_{D-P}$ in the two-dimensional subspace 
spanned by $\ket{G_{\text{N\'{e}el}}}\otimes\ket{1_P}$ and 
$\ket{E_{\text{N\'{e}el}}}\otimes\ket{0_P}$ results in two polariton states 
with energy splitting 
\begin{align}
 g_{\text{eff}}[\text{N\'{e}el}] = 2 g \sqrt{2 L}\,. 
 \label{gneel}
\end{align}
This value needs to be compared to 
$g_{\text{eff}}$ in Eq.~(\ref{geff}). We find that  $g_{\text{eff}}$ is 
larger than $g_{\text{eff}}[\text{N\'{e}el}]$ by a factor of 
$\sqrt{2\log 2}\approx 1.18$, i.e., 
$g_{\text{eff}}/g_{\text{N\'{e}el}}[\text{N\'{e}el}]=\sqrt{2\log 2}$. The reason for this  is that $ \ket{G_{\text{N\'{e}el}}}$ is not the true  ground state of the electronic system, which is  an entangled superposition of  Wannier states. 
It follows that the correlations in the true ground state of the electronic 
system enhance the polariton splitting by about $18\%$. 

Next we compare our results to those obtained for $L$ independent two-level atoms with transition energy $U$  that interact resonantly with a single cavity mode. This model is a special case of the so-called 
Tavis-Cummings~\cite{tavis:68,tavis:69} or Dicke~\cite{dicke2:54} model, and in the following we refer to it as the Dicke model. A brief description of 
the Dicke Hamiltonian is given in Appendix~\ref{dicke}.
The manifold with zero excitations  has only one non-degenerate ground state where  the cavity is in the vacuum state and all atoms are in the ground state. 
The manifold with one excitation contains two states that are split by [see Appendix~\ref{dicke}] 
\begin{align}
 g_{\text{eff}}[\text{Dicke}] = 2 g \sqrt{L} \,,
 \label{gdicke}
\end{align}
which is smaller than 
$g_{\text{eff}}[\text{N\'{e}el}]$ in Eq.~(\ref{gneel}) by a factor 
of $\sqrt{2}$.  
This difference can be attributed to the different nature of the 
light-matter interaction for atoms and electrons: The cavity field couples to the atomic density in the Dicke model, whereas the light-matter coupling 
in the electronic system is proportional to the  current operator.  
Starting from $\ket{G_{\text{N\'{e}el}}}$ the operator $\hat{H}_{D-P}$ can 
create a doublon with the holon either to the left or two the right, 
giving rise to $2L$ possible states as discussed above. 
On the contrary, the corresponding Hamiltonian for the atoms can only 
locally excite one atom at site $k$, and there are only $L$ different 
states. Taking into account the normalisation of the corresponding states gives rise to collective coupling strengths proportional 
to $\sqrt{2L}$ and $\sqrt{L}$ in the case of electrons and atoms, respectively. 
\section{Experimental realization \label{exp}}
To discuss the experimental observation of the collectively enhanced 
light-matter coupling in our system 
we consider $\text{ET-F}_2\text{TCNQ}$~\cite{hasegawa:97,hasegawa:00,wall:11,mitrano:14},  which is a generic example of a one-dimensional Mott insulator where $U \gg t$. 
In order to observe the splitting of the optical conductivity spectrum shown in Fig.~\ref{fig4} we require 
$g_{\text{eff}}/t  \gtrsim 3$. In addition, the photon-doublon coupling  must be much faster than the 
cavity decay rate $\kappa$, i.e.,  $g_{\text{eff}}\gg\hbar\kappa$. 
In the case of $\text{ET-F}_2\text{TCNQ}$~\cite{hasegawa:97} we find  
 $g_{\text{eff}}/t \approx 6.4$. 
 It follows that  the two  branches in the optical conductivity should be 
 clearly visible, and their energy splitting can be as large as 
$g_{\text{eff}}\approx 250\text{meV}$. 
 Even larger values of $g_{\text{eff}}/t$ are possible in materials with a smaller Mott gap or smaller unit cells. 
The condition $g_{\text{eff}}\gg\hbar\kappa$ is also fulfilled in $\text{ET-F}_2\text{TCNQ}$ where $t/\hbar\approx 2\pi \times 10\text{THz}$~\cite{hasegawa:97,hasegawa:00}, 
which is at least two orders of magnitude larger than  cavity decay rates of lossy microcavities with frequencies in the low THz regime~\cite{keller:17}. 

Finally we address the finite lifetime $\tau_D$ of doublon excitations which 
increases exponentially with $U/t$~\cite{strohmaier:10}. 
The experimentally measured value for $\text{ET-F}_2\text{TCNQ}$  
at ambient pressure is $\tau_D\approx 0.5\text{ps}$~\cite{mitrano:14}, 
which corresponds 
to a decay rate of $\kappa_D\approx 0.2 t/\hbar$. This decay rate is 
smaller than the artificial broadening introduced  in the numerical evaluation  of Eq.~(\ref{OCsimple}), where each term of the sum was  broadened with a Lorentzian of width $0.5t/\hbar$. We thus conclude that 
the finite lifetime of doublons does not hinder the observation of the 
two peaks in the optical conductivity. 
\section{Summary \label{summary}} 
We have shown that the resonant coupling between strongly 
correlated electrons and a single-mode cavity results in the formation 
of Mott polaritons. The manifold with one excitation exhibits a dense 
spectrum of polariton branches which can be derived from the 
eigenstates in the zero excitation manifold. 
At half filling the effective Hamiltonian 
in the manifold with zero excitations is an isotropic Heisenberg chain. 
Each eigenstate with non-zero eigenvalue $\mc{E}_j<0$ gives rise to two polariton branches, and the magnitude of their energy splitting is proportional to $\sqrt{-\mc{E}_j}$. 
The two branches with the largest energy splitting are thus associated 
with the ground state of the isotropic Heisenberg chain, and their 
energy splitting $g_{\text{eff}}$ is proportional to $\sqrt{2L}$, where $L$ is the number of electronic sites.

An approximate derivation for $g_{\text{eff}}$  in Sec.~\ref{discuss}  
illustrates that quantum correlations in the ground state result in 
an enhancement of the polariton splitting by $18\%$. 
Furthermore, $g_{\text{eff}}\propto \sqrt{2L}$ is a direct consequence 
of the fact that the electron-photon interaction is mediated by the 
current operator. The absorption of a photon is associated with an 
electronic hopping process creating a holon-doublon pair where the 
doublon is either to the right or the left of the holon. This two-fold 
excitation pathway is in 
contrast to atomic systems where the atomic density couples to the 
cavity field, allowing only for one local excitation when absorbing a 
photon. The collective polariton splitting for $L$ independent two-level atoms and in the manifold with 
one excitation  is consequently smaller by a factor of $\sqrt{2}$ compared to our electronic system. 

We find that the collectively enhanced polariton splitting is directly observable in the optical conductivity, which features 
two peaks separated by $g_{\text{eff}}\propto\sqrt{2L}$. If the material fills the 
whole mode volume of the cavity, the magnitude of the splitting is 
independent of the mode volume and just depends on $1/\sqrt{v_{\text{uc}}}$, 
where $v_{\text{uc}}$ is the volume of the unit cell of the crystal. 
As a generic example of a one-dimensional Mott insulator we consider 
$\text{ET-F}_2\text{TCNQ}$, and find that its unit cell is small enough 
such that the splitting of the optical conductivity signal exceeds the 
width of the first Hubbard band. The optical conductivity thus carries 
a clear signature of the collective electron-photon coupling. 

We emphasise that $g_{\text{eff}}\propto1/\sqrt{v_{\text{uc}}}$ together 
with the small unit cells  in 
solid state materials can result in macroscopically large 
polariton splittings $g_{\text{eff}}$. 
In the case of $\text{ET-F}_2\text{TCNQ}$, we find 
$g_{\text{eff}}\approx 250 \text{meV}$, which is several orders of 
magnitude larger than what has been achieved in atomic 
systems~\cite{raizen:89,thompson:92,baumann:10}. 
In addition, we note that the near-resonant electron-photon coupling 
described in this work is much larger than the effects described 
in~\cite{kiffner:19,kiffner:19b}, which are mediated by virtual, 
second-order electron-photon interactions. 

%
%
In this paper we focussed on the resonant electron-photon coupling 
in the manifold with one excitation. An intriguing prospect for 
future studies is to investigate higher-excited manifolds where 
different sub-manifolds are resonantly coupled via the electron-photon 
interaction as indicated in Fig.~\ref{fig2}.  
Since  the electron-photon interaction increases with the 
number of photons $j$ as $\sqrt{j}$, the energy spectrum 
is anharmonic. Like in atomic systems~\cite{schuster:08} this feature results in giant photon nonlinearities and further amplifies 
the intrinsically large optical nonlinearity of Mott insulators~\cite{kishida:00}.  
Furthermore, the physics in  manifolds with a large number 
of excitations will be fundamentally different from the Dicke model. 
The reason is  that 
the maximal number of atomic excitations within the Dicke model is $L$, 
but at most $L/2$ doublons can be created in the electronic system.
A further intriguing avenue for future studies is the investigation 
of higher-dimensional systems. For example, 
the electron-cavity interaction in higher-dimensional systems 
can be tuned via the relative orientation between the crystal and the cavity polarization vector~\cite{kiffner:19,kiffner:19b}. 
In $k$-dimensional systems where the cavity couples to all spatial directions one expects  
$g_{\text{eff}}\propto\sqrt{2k}$ due to the additional excitation 
pathways to nearest-neighbour sites, and thus a further enhancement of 
the effective coupling strength.
\begin{acknowledgments}
MK and DJ acknowledge financial support  from the National Research Foundation, Prime Minister’s Office, Singapore, and
the Ministry of Education, Singapore, under the Research
Centres of Excellence program. DJ,FS and JC acknowledge funding from the 
European Research Council under the European Unionʼs Seventh Framework Programme 
(FP7/2007-2013)/ERC Grant Agreement no. 319286, Q-MAC. DJ acknowledges 
funding from  EPSRC grant no. EP/P009565/1. MK, AA and DJ thank
Andrea Cavalleri for discussions. 
\end{acknowledgments}
%
%
\appendix
\section{Isotropic Heisenberg model \label{hbc}}
The effective Hamiltonian $\hat{H}_{\text{S}}$  in Eq.~(\ref{Hex}) can be cast into the form~\cite{essler:05}
\begin{align}
 \hat{H}_{\text{S}} = J_c\, \sum\limits_{ \langle jk\rangle}
 \left(\mf{\hat{S}}_j\cdot\mf{\hat{S}}_k - \frac{\hat{n}_j \hat{n}_k}{4}\right)\hat{\mc{P}}_0^D\,,
\end{align}
where $\hat{n}_j=\hat{c}_{j,\uparrow}^{\dagger}\hat{c}_{j,\uparrow} +
\hat{c}_{j,\downarrow}^{\dagger}\hat{c}_{j,\downarrow}$ is the number operator at site $j$ and the components of the local spin operator 
$\mf{\hat{S}_j}=(\hat{S}_x^j,\hat{S}_y^j,\hat{S}_z^j)$ are defined as 
\begin{subequations}
 \begin{align}
  \hat{S}_x^j = \frac{1}{2}\left(
\hat{c}_{j,\uparrow}^{\dagger}\hat{c}_{j,\downarrow} +   
\hat{c}_{j,\downarrow}^{\dagger}\hat{c}_{j,\uparrow} 
  \right) \,, \\ 
  \hat{S}_y^j = \frac{\imag}{2}\left(
  \hat{c}_{j,\downarrow}^{\dagger}\hat{c}_{j,\uparrow} -
\hat{c}_{j,\uparrow}^{\dagger}\hat{c}_{j,\downarrow}
  \right) \,, \\ 
  \hat{S}_z^j = \frac{1}{2}\left(
  \hat{c}_{j,\uparrow}^{\dagger}\hat{c}_{j,\uparrow} -
\hat{c}_{j,\downarrow}^{\dagger}\hat{c}_{j,\downarrow}
  \right) \,. 
 \end{align}
\end{subequations}
At half filling every site is occupied precisely by one electron, and thus 
\begin{align}
 \hat{H}_{\text{S}} = J_c\, \sum\limits_{ \langle jk\rangle}
 \left(\mf{\hat{S}}_j\cdot\mf{\hat{S}}_k - \frac{1}{4}\right)\hat{\mc{P}}_0^D 
\end{align}
is an isotropic spin-1/2 Heisenberg chain with exchange coupling  $J_c$.
\section{Effective Hamiltonian in $\hat{\mc{P}}_1$ \label{sx}}
An approximate effective Hamiltonian in the manifold $\hat{\mc{P}}_1$ 
that only takes into account the perturbation $\hat{H}_1$ in first order  
and for $\Omega=U$ is~\cite{tannoudji:api,essler:05} 
 \begin{align}
  \hat{H}_{\text{eff}}^{(1)} = 
  U \hat{\mc{P}}_1
  + \hat{\mc{P}}_1 \hat{H}_1 \hat{\mc{P}}_1 
  \,.
  \label{exD}
 \end{align}
The eigenvalues of $\hat{H}_{\text{eff}}^{(1)}$ will coincide with 
the eigenvalues of the full Hamiltonian $\hat{H}$ in $\hat{\mc{P}}_1$ 
in the limit $U\gg g,t$ where higher-order terms in $\hat{H}_1$ 
are negligible.  
The first term in Eq.~(\ref{exD}) represents the 
unperturbed energy of the manifold  with one excitation, which can be either 
a photon with energy $\Omega=U$ or a doublon. The second term in Eq.~(\ref{exD}) can be written as 
\begin{align}
 \hat{\mc{P}}_1 \hat{H}_1 \hat{\mc{P}}_1 = \mc{R}_c \hat{\mc{P}}_1^{(0)}\hat{T}\hat{\mc{P}}_1^{(0)} 
 +  g \hat{\mc{P}}_1 
 \left[(\hat{a}+\hat{a}^{\dagger}) \hat{\mc{J}} \right]\hat{\mc{P}}_1 \,,
 \label{p1h1}
\end{align}
where we used   
$\hat{\mc{P}}_1^{(1)}\hat{T}\hat{\mc{P}}_1^{(1)} = 0$ at half filling. 
The second term in Eq.~(\ref{p1h1}) is 
\begin{align}
g \hat{\mc{P}}_1 
 \left[(\hat{a}+\hat{a}^{\dagger}) \hat{\mc{J}} \right]\hat{\mc{P}}_1   = \hat{H}_{D-P} \,,
 \label{heq}
\end{align}
where $\hat{H}_{D-P}$ is defined in Eq.~(\ref{Hdp}). Combining 
Eqs.~(\ref{p1h1}),~(\ref{heq}) and Eq.~(\ref{exD})  
shows that the expression for $\hat{H}_{\text{eff}}^{(1)}$ in 
Eq.~(\ref{exD}) is the same as  Eq.~(\ref{oneEx}). 
%

\begin{figure}[t!]
\begin{center}
\includegraphics[width=\columnwidth]{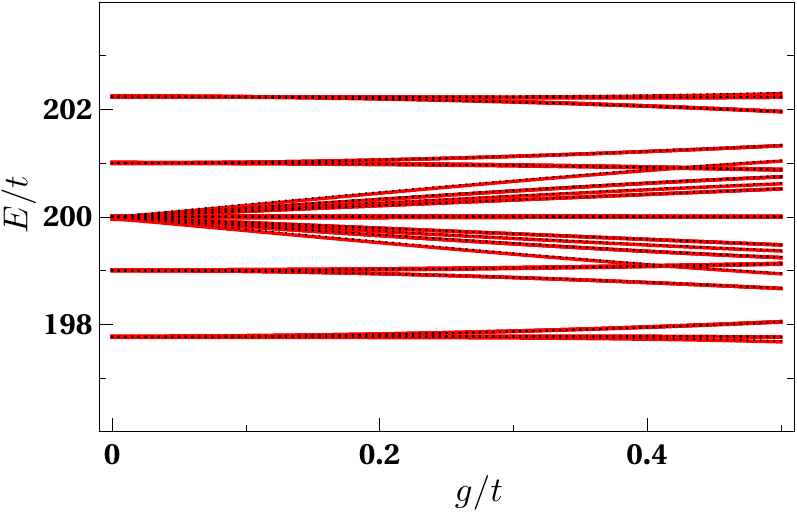}
\end{center}
\caption{\label{fig4}
(Color online) Comparison of the eigenenergies $E$ of the system Hamiltonian $\hat{H}$  in the  manifold with one excitation and 
 $\hat{H}_{\text{eff}}^{(1)}$ defined in Eq.~(\ref{oneEx}). 
 We consider  a system with $L=4$ sites at half filling and show $E$ as a function of the cavity coupling $g$. 
The eigenvalues corresponding to $\hat{H}$ ($\hat{H}_{\text{eff}}^{(1)}$) are shown by red solid (black dotted) lines. 
The exact diagonalization calculations take into account photon states 
$\ket{j_P}$ with $j \in \{0,1,2,3,4\}$, and we set $U=\Omega = 200 t$. 
}
\end{figure} 
%
A comparison of the eigenvalues of the effective Hamiltonian $\hat{H}_{\text{eff}}^{(1)}$  and the system Hamiltonian $\hat{H}$ is shown in Fig.~\ref{fig4} 
for a system with $L=4$ sites at half filling and as a function of the cavity coupling $g$. The eigenvalues are in very good agreement for large values of 
$U=\Omega$, and their differences are of the order of higher-order  corrections  $\mc{O}(g^2/U,t^2/U,gt/U)$ that are neglected in Eq.~(\ref{exD}). 
\section{Spectrum of $\hat{H}_{D-P}$ \label{SpecHdp}}
Here we investigate the spectrum of the Hamiltonian $\hat{H}_{D-P}$ in  
Eq.~(\ref{Hdp}). To this end we consider matrix elements of $\hat{H}_{D-P}$ 
between states in the $\hat{\mc{P}}_1$ manifold. Note that $\hat{H}_{D-P}$ 
can only couple states with one photon and no doublon to states with 
no photon and one doublon. In a first step, we construct 
electronic states $\ket{e_{j}}$ with exactly one doublon from 
the  eigenstates  $\ket{g_{j}}$ of $\hat{H}_{\text{S}}$  
with non-zero eigenvalue $\mc{E}_j<0$, 
\begin{align}
 \ket{e_{j}} = \frac{1}{2 \epsilon_{j}} \hat{\mc{D}}^{\dagger}\ket{g_{j}}\,,
 \label{ej}
\end{align}
where $\hat{\mc{D}}$ and $\epsilon_{j}$ are defined in Eqs.~(\ref{Dop}) 
and~(\ref{epsj}), respectively. 
At half filling it is straightforward to prove the operator identity 
\begin{subequations}
 \label{iden}
\begin{align}
 \hat{\mc{D}} \hat{\mc{D}}^{\dagger} &
 = -\frac{4}{J_c} \hat{H}_{\text{S}}\\
  & = 4 \hat{\mc{P}}_0^D \left(\sum\limits_{\langle kl \rangle} \hat{b}_{kl}^{\dagger}\hat{b}_{kl}\right) \hat{\mc{P}}_0^D\,,
\end{align} 
\end{subequations}

and hence the states $\ket{e_{j}}$ are orthonormal, 
\begin{align}
  \braket{e_{i}}{e_{j}} =\frac{1}{4 \epsilon_{i}\epsilon_{j}} \bra{g_{i}} \hat{\mc{D}} \hat{\mc{D}}^{\dagger} \ket{g_{j}} = \delta_{ij}\,.
 \label{ortho}
\end{align}
Next we define the following states in $\hat{\mc{P}}_1$, 
\begin{subequations}
\label{PolStates}
 \begin{align}
 \ket{\psi_{P}^{(j)}} & =  \ket{g_{j}}\otimes\ket{1_P}\,,\\
 \ket{\psi_{D}^{(j)}} & = \ket{e_{j}}\otimes\ket{0_P}\,.
 \end{align}
\end{subequations}
The  matrix elements of $\hat{H}_{D-P}$ with respect to these states 
can be found via Eq.~(\ref{ortho}) and are given by 
\begin{align}
 \bra{\psi_{D}^{(i)}}\hat{H}_{D-P}\ket{\psi_{P}^{(j)}} = 2g\epsilon_{i}\,\delta_{ij}\,.
 \label{nzm}
\end{align}
It follows that the matrix representation of $\hat{H}_{D-P}$ 
reduces to a simple $2\times2$ block diagonal form in the states defined in Eq.~(\ref{PolStates}), and diagonalizing these blocks leads to the 
polariton states in Eq.~(\ref{polariton}). 

It remains to show that the matrix elements in Eq.~(\ref{nzm}) and their 
complex conjugates are the only non-zero matrix elements of 
$\hat{H}_{D-P}$ in $\hat{\mc{P}}_1$. This can be understood as follows. 
First, we consider states $\ket{\psi_D^{(k)}}_{\perp}$ that complement the states 
$\ket{\psi_D^{(j)}}$ to an orthonormal basis in $\hat{\mc{P}}_1^{(0)}$. 
Since 
\begin{align}
\hat{H}_{D-P}\ket{\psi_{P}^{(j)}}\propto\ket{\psi_{D}^{(j)}}
\end{align}
according to Eqs.~(\ref{Hdp}) and~(\ref{ej}), we find
\begin{align} 
 {}_{\perp}\bra{\psi_D^{(k)}}\hat{H}_{D-P}\ket{\psi_{P}^{(j)}} = 0
 \label{nzm2}
\end{align}
for all  values of $j$ and $k$. 
Second, we consider the eigenstates $\ket{g_l^0}$ of $\hat{H}_{\text{S}}$ 
with eigenvalue zero. The states 
\begin{align}
 \ket{\psi_P^{(l)}}_{\perp} = \ket{g_l^0}\otimes\ket{1_P}
\end{align}
complement the states $\ket{\psi_P^{(l)}}$ to a basis in 
$\hat{\mc{P}}_1^{(1)}$. 
According to Eq.~(\ref{iden}) we have 
\begin{align}
 \bra{g_{l}^0} \hat{\mc{D}} \hat{\mc{D}}^{\dagger} \ket{g_{l}^0} =0\,,
\end{align}
and thus $\hat{\mc{D}}^{\dagger} \ket{g_{l}^0} =0$. It follows that all matrix elements of $\hat{H}_{D-P}$ involving $ \ket{\psi_P^{(l)}}_{\perp}$ vanish,
\begin{subequations}
 \label{nzm3}
\begin{align} 
 \bra{\psi_{D}^{(j)}}\hat{H}_{D-P} \ket{\psi_P^{(l)}}_{\perp} 
 = &0\,,\\
 {}_{\perp}\bra{\psi_D^{(k)}}\hat{H}_{D-P} \ket{\psi_P^{(l)}}_{\perp}
 = &0\,,
\end{align}
\end{subequations}
which concludes our proof of the spectrum of $\hat{H}_{D-P}$.

Finally, we note that  the polariton states $\ket{\psi_{D-P}^{(j)}}_{\pm}$ are not coupled by  the kinetic energy term 
$\hat{\mc{P}}_1^{(0)}\hat{T}\hat{\mc{P}}_1^{(0)}$, 
\begin{align}
 &  \bra{\psi_{D-P}^{(i)}}_{\pm}\hat{\mc{P}}_1^{(0)}\hat{T}\hat{\mc{P}}_
 1^{(0)}\ket{\psi_{D-P}^{(j)}}_{\pm} 
 \propto  \bra{e_{i}}\hat{\mc{P}}_1^{(0)}\hat{T}\ket{e_{j}} =0\,. 
 \label{zeroM}
\end{align}
The second equality in Eq.~(\ref{zeroM}) follows from 
the fact that in  $\ket{e_{i}}$, the doubly occupied site has an adjacent empty site to its right or left. 
On the other hand,  $\hat{\mc{P}}_1^{(0)}\hat{T}\ket{e_{j}}$ 
is either zero or describes a state where the doublon and the holon are separated by a singly occupied site, and hence this state is orthogonal 
to $\ket{e_i}$. 

Note that $\hat{\mc{P}}_1^{(0)}\hat{T}\hat{\mc{P}}_1^{(0)}$ has non-zero 
matrix elements between the states $\ket{\psi_D^{(k)}}_{\perp}$ that 
form the quasi-continuum of the first Hubbard band. Furthermore, 
 $\hat{\mc{P}}_1^{(0)}\hat{T}\hat{\mc{P}}_1^{(0)}$ couples 
$\ket{\psi_D^{(k)}}_{\perp}$ to 
the polariton states $\ket{\psi_{D-P}^{(j)}}_{\pm}$ via 
the states $\ket{\psi_D^{(j)}}$. This coupling leads to a broadening of the 
polariton states but  becomes less effective if the polariton 
splitting exceeds the tunneling amplitude $t$. We thus  
expect the resonances in the optical conductivity to become sharper when the collective coupling increases. 
\section{Dicke model \label{dicke}}
The Tavis-Cummings or Dicke Hamiltonian for a system of $L$ independent  
two-level atoms interacting with a single cavity mode and in 
rotating-wave approximation is given 
by~\cite{dicke2:54,tavis:68,tavis:69,garraway:11}
\begin{align}
 H_{\text{Dicke}} = \Omega \hat{a}^{\dagger} \hat{a} + U S_z + g\left(a^{\dagger}S_- + a S_+\right)\,,
 \label{Hdicke}
\end{align}
where   $g$ is the light-matter coupling constant. The collective atomic operators are defined as 
\begin{subequations}
\begin{align}
 S_z & = \frac{1}{2}\sum\limits_{k=1}^L \left(
 \ket{e_k}\bra{e_k} - \ket{g_k}\bra{g_k}
 \right) \,,\\
 S_+ & = \sum\limits_{k=1}^L \ket{e_k}\bra{g_k}\,,\\
 S_- & = S_+^{\dagger}= \sum\limits_{k=1}^L \ket{g_k}\bra{e_k} \,,
\end{align}
\end{subequations}
where $\ket{g_k}$ ($\ket{e_k}$) denotes the ground (excited) 
state for the $k$th atom. 
There are two eigenstates of $H_{\text{Dicke}}$ in the subspace of one excitation, and their energy difference for $\Omega=U$ 
is 
$g_{\text{eff}}[\text{Dicke}]$
defined in Eq.~(\ref{gdicke}).

%
%

%
\end{document}